\documentclass[fleqn,usenatbib]{mnras}

\usepackage{mathptmx}
\usepackage{rotating}

\usepackage[T1]{fontenc}

\DeclareRobustCommand{\VAN}[3]{#2}
\let\VANthebibliography\thebibliography
\def\thebibliography{\DeclareRobustCommand{\VAN}[3]{##3}\VANthebibliography}

\usepackage{amsmath}	% Advanced maths commands
\usepackage{amssymb}	% Extra maths symbols
\usepackage{pdflscape}
\usepackage{graphicx} % Including figure files
\usepackage{soul}
\usepackage{multirow}

\makeatletter
\newcount\SOUL@minus %% <- without this line the page number would be reset to 0
\makeatother

\title[Morphological classification of radio galaxies using convolutional NN and data augmentation]{Morphological classification of compact and extended radio galaxies using convolutional neural networks and data augmentation techniques}

\author[V. Maslej-Kre\v{s}\v{n}\'akov\'a, K. El Bouchefry and P. Butka]{
Viera Maslej-Kre\v{s}\v{n}\'akov\'a$^{1}$, Khadija El Bouchefry$^{2}$ and Peter Butka$^{1}$\thanks{E-mail: peter.butka@tuke.sk}
\\
$^{1}$Department of Cybernetics and Artificial Intelligence, Faculty of Electrical Engineering and Computer Science, Technical University of Košice, Slovakia\\
$^{2}$South African Radio Astronomy Observatory, Johannesburg, South Africa \\
}

\date{Accepted 2021 May 12. Received 2021 May 11; in original form 2020 December 10}

\pubyear{2021}

\begin{document}
\label{firstpage}
\pagerange{\pageref{firstpage}--\pageref{lastpage}}
\maketitle

\begin{abstract}
 %Next generation radio surveys and observatories such as the Square Kilometre Array (SKA) will produce massive amounts of radio imaging data. Traditional classification based on human visual inspection or using classical techniques of large numbers of sources with different morphologies is almost impossible.  Machine learning techniques have been increasingly used in astronomical applications and have proven to  successfully  classify objects in image data with high accuracy. Morphological classification of radio galaxies into different classes provides insights into central black holes and their accretion modes, interactions between radio galaxies, and can also be used  to probe dark matter, etc. 
 
 Machine learning techniques have been increasingly used in astronomical applications and have proven to  successfully  classify objects in image data with high accuracy. The current work uses archival data from the Faint Images of the Radio Sky at Twenty Centimeters (FIRST) to classify radio galaxies into four classes: Fanaroff-Riley Class I (FRI), Fanaroff-Riley Class II (FRII), Bent-Tailed (BENT), and Compact (COMPT). The  model presented in this work is based on Convolutional Neural Networks (CNNs). The proposed architecture comprises three parallel blocks of convolutional layers combined and processed for final classification by two feed-forward layers. Our model classified selected classes of radio galaxy sources on an independent testing subset with an average of 96\% for precision, recall, and F1 score. The best selected augmentation techniques were rotations, horizontal or vertical flips, and increase of brightness. Shifts, zoom and decrease of brightness worsened the performance of the model. The current results show that model developed in this work is able to identify different morphological classes of radio galaxies with a high efficiency and performance.
%This is a simple template for authors to write new MNRAS papers.
%The abstract should briefly describe the aims, methods, and main results of the paper.
%It should be a single paragraph not more than 250 words (200 words for Letters).
%No references should appear in the abstract.

\end{abstract}


%%%%%%%%%%%%%%%%%%%%%%%%%%%%%%%%%%%%%%%%%%%%%%%%%%

%%%%%%%%%%%%%%%%% BODY OF PAPER %%%%%%%%%%%%%%%%%%

\section{Introduction}
\label{intro}
The new upcoming next generation of radio observatories such as the Australian SKA Pathfinder \cite[ASKAP;][]{Johnston2007,Johnston2008,Johnston2009}, The Expanded Karl G. Jansky Very Large Array \cite[EVLA;][]{Perley2011}, and the Square Kilometre Array \cite[SKA;][]{Braun2015, Quinn2015, dewdney2009square}  will produce large amounts of radio imaging data. For example, MeerKCLASS (an individual MeerKat survey) is expected to detect more than 200 000 radio sources including HI radio galaxies, star-forming galaxies, and other types of radio sources \citep{Santos2017}, and the Evolutionary Map of the Universe \cite[EMU;][]{Hopkins2015} survey  conducted by ASKAP is expected to find about 70 million radio sources \citep{Norris2011}. Processing and analysing this massive amount of data in a classical manner is almost impossible. The only real possibility to automatically and rapidly process and analyse streams of imaging data (including morphological classification) is in the development  of new automatic algorithms. 

Radio galaxies can be classified either as compact or extended radio morphologies \citep{Miraghaei2017}. A compact radio galaxy (COMPT) is visually simple, best described as a single non-diffuse object in radio image. Most of the radio galaxy sources detected at 1.4GHz are compact \citep{banfield2015radio}. Extended radio galaxies have been broadly split into two groups known as FRI and FRII based on Fanaroff-Riley (FR) scheme \citep{fanaroff1974morphology}. The FR scheme is based on the ratio $R_{RF}$ of  the distance between the regions of the highest   surface brightness on the extended components of a source in opposite directions and the largest angular extent of the source. Radio sources with $R_{RF} < 0.5$ are classified as FRI, and radio sources with $R_{RF}> 0.5$ are classified as FRII. Moreover, FRI  radio galaxies have bright regions along the jets and core, and reside in moderately rich cluster environments \citep{Hill1991}. In contrast,  FRII radio galaxies contain strong radio emissions in their extremities lobes, as well as more collimated jets. FRI and FRII do live in different environment (on average), while FRII at higher redshifts live in rich groups \citep{Zirbel1997}. There is also a sharp division in radio luminosity between the two classes at $ L_{178\, MHz} \approx 2\times  10^{25} W\, Hz^{-1} \, sr^{-1}$,  with FRIs below and FRIIs above this threshold \citep{fanaroff1974morphology, Owen1994}. Detailed discussion  on FR dichotomy can be found in \cite{Saripalli2012}. While the majority of radio sources population is comprised of FRI and FRII morphologies,  more morphological subclasses have been identified such as Narrow-Angle Tail sources (NAT, \citet{Rudnick1976, giacintucci2009tailed}), Wide-Angle Tail sources (WAT, \citet{Owen1976, giacintucci2009tailed}), Double-Double Radio Galaxies \citep[DDRG; ][]{Schoenmakers2001, Saikia2006}, X-shaped radio galaxies (XRG, \citet{Leahy1992, Cheung2007, Yang2019, Bhukta2020}), ring like radio galaxies \citep{Proctor2011}, and  Hybrid Morphology Radio Sources \citep[HYMORS;][]{Gopal-Krishna2000, banfield2015radio, Kapinska2017}. Classification of radio galaxies into different morphologies provides information on the formation and evolution of galaxies, and also allows us to understand their cosmic environment  \citep{helfand2015last}. Studying their centres or cores which are believed to be powered by  super-massive black holes allows  us to probe the dynamics of jets, their evolution, and their connection to the host galaxy \citep{Makhathini2015}.

The morphological classes of radio galaxies were traditionally identified using visual examination, with examples of catalogs like the NRAO VLA Sky Survey  \citep[NVSS; ][]{Condon1998},  Sydney University Molonglo Sky Survey \citep[SUMSS;][]{sumss} and the Faint Images of the Radio Sky at Twenty Centimeters \citep[FIRST;][]{first} . The citizen science project Radio Galaxy Zoo \citep[RGZ;][]{banfield2015radio, Wu2019}provided more than two million of annotations. However, even with a large number of annotations, it is not feasible to classify objects manually using data  collected by the upcoming telescope arrays like the SKA, which will produce a massive amount of data every day. SKA is expected to produce about 3ZB of raw data \cite[150 TB per day;][]{Garofalo2017} including images, catalog and redshifts. Typical image size from MeerKAT telescope for example is about 11.13 TB \citep{3}. Therefore, there is a need for automated methods and algorithms for processing and analysis \citep{Norris2017}. Another aspect to consider when handling massive amount of data is the computational complexity, which also affects data processing costs.

In recent years, machine learning algorithms successfully provide new opportunities and tools for processing data from astronomical surveys \citep[e.g.][]{2,3}. Deep learning methods, specifically Convolutional Neural Networks (CNNs, \cite{lecun2015deep}) became widespread technique in astronomy and astrophysics for the detection and classification of astronomical objects. CNNs have been successfully applied in the following areas: detection and classification of galaxies \citep{Ackermann2018, galaxy2, galaxy4}, detection of gravitational waves and glitches \citep{deep1, deep2, glitches1, glitches2}, classification of gravitational lensing objects \citep{lens1, lens2}, or even classification of objects according to their  light curves \citep{light_curve2}. However, the classification of radio sources images using deep learning was first presented by \citet{3}, who demonstrated that the neural networks can accurately recognise different classes of radio galaxies. The authors used archival data from FIRST to classify radio galaxies into FRI, FRII and BENT radio morphologies with a class accuracy of 91 per cent, 75 per cent, and 95 per cent, respectively. \citet{2} have extended the latter work to include compact sources using similar data samples. \citet{2} developed a CNN model with three layers and some augmentation methods  to classify  FIRST radio sources into FRI, FRII, BENT, and COMPT morphologies achieving  class accuracies of 98 percent, 100 per cent, 98 per cent, and 93 per cent, respectively. Another application of CNNs to radio galaxy morphology was conducted by \citet{Lukic_2018}, who developed a model based on three convolutional layers and two dense layers, with an accuracy of 92 per cent. \cite{1} used transfer learning for the morphological classification of radio galaxies. They applied a convolutional neural network with 13 layers on FIRST and NVSS catalogs with 89 per cent accuracy, but only FRI and FRII classes were used. 

In all of the mentioned works, the authors applied augmentation techniques with a different setup of split and augmentation steps. In \cite{3,1}, and \cite{Lukic_2018}, the authors have ensured that they separated samples in their testing sets from the samples in training sets before applying any data augmentation technique on the training part. In this case, the test set is independent from all augmentations, and thus their results represent the full model generalization ability. Table~ \ref{table_sota} shows more details on their results, with precision, recall, and F1 score for every class. On the other hand, the testing subset by authors in \cite{2} first rotated original images and then split them into training and test set. Therefore, there are cases where the original image can be in both subsets but is always in different rotations.  Table~ \ref{table_sota2} shows more results of this work. While the second approach slightly reduces generalization ability, it still provides some confidence in results because samples are independent according to the rotation. Because previous works provide both approaches, we decided to create models and compare experiments for both setups – independent split and split on already rotated original images.

\begin{table}
\centering
\caption{The results of classification provided in related work with the setup based on their independent testing subset. Support represents the number of test samples for each class, with sum of them as total.}
\label{table_sota}
\begin{tabular}{lcccr}
\hline
                 & \textbf{Precision} & \textbf{Recall} & \textbf{F1 score} &  \textbf{Support}\\ \hline
                 \multicolumn{5}{l}{Results from \cite{3}}                                                       \\ \hline
BENT             & 0.95              & 0.79                & 0.87              & 77             \\ 
FRI              & 0.91              & 0.91                & 0.91              & 53             \\ 
FRII             & 0.75              & 0.91                & 0.83             & 57             \\ \hline
\textbf{average}   & \textbf{0.88}     & \textbf{0.86}       & \textbf{0.86}             & \textbf{total: 187}           \\ \hline

\multicolumn{5}{l}{Results from \cite{Lukic_2018}}                                                       \\ \hline
COMPT            & 0.97              & 0.97                & 0.97              & -           \\ 
BENT             & 0.93              & 0.95                & 0.94              & -           \\ 
FRI              & 0.91              & 0.88                & 0.89             & -           \\ 
FRII             & 0.95              & 0.96                & 0.95             & -          \\ \hline
\textbf{average}  & \textbf{0.94}     & \textbf{0.94}       & \textbf{0.94}             & \textbf{}         \\ \hline

\multicolumn{5}{l}{Results from \cite{1}}                                                       \\ \hline
FRI              & 0.95              & 0.85                & 0.90              & 80             \\ 
FRII             & 0.83              & 0.94                & 0.88              & 117            \\ \hline
\textbf{average} & \textbf{0.89}     & \textbf{0.89}       & \textbf{0.89}             & \textbf{total: 197}            \\ \hline
\end{tabular}
\end{table}

\begin{table}
\centering
\caption{The classification results provided in related work with the setup based on a split of already rotated images. Support represents the number of test samples for each class, with sum of them as total.}
\label{table_sota2}
\begin{tabular}{lcccr}
\hline
                 & \textbf{Precision} & \textbf{Recall} & \textbf{F1 score} &  \textbf{Support}\\ \hline
\multicolumn{5}{l}{Results from \cite{2}}                                                       \\ \hline
COMPT            & 0.98              & 0.98                & 0.98              & 1000           \\ 
BENT             & 0.96              & 0.98                & 0.97              & 1000           \\ 
FRI              & 0.98              & 1.00                & 0.99              & 1000           \\ 
FRII             & 0.96              & 0.93                & 0.95              & 1000           \\ \hline
\textbf{average}  & \textbf{0.97}     & \textbf{0.97}       & \textbf{0.97}             & \textbf{total: 4000}       \\
\hline
\end{tabular}
\end{table}

Deep learning models are specific with their need for an adequate amount of input data in order to extract necessary features during training \citep{Goodfellow-et-al-2016}. Therefore, if we start with a smaller or limited dataset of labeled data, we need to enhance the learning process with some augmentation techniques. Such techniques can enlarge the input dataset, especially in imaging data, using different transformations on the original data \citep{shorten2019survey}. 

The purpose of the current work is to automate the morphological classification of radio sources by developing a CNN classifier that can classify large amounts of radio galaxies with high accuracy. In order to do this we developed  our CNN classifier based on the combination of three neural network blocks  followed by two fully connected layers for final classification (i.e. Figure \ref{fig:model}). As a data set, we used  four morphological classes: FRI, FRII, BENT, and COMPT radio galaxies. This work, also emphasized the application of augmentation techniques on the input dataset for the training of our convolutional neural network architecture. Our classifier provides relatively simple architecture with similar or better performance to previous works.

This paper is organized as follows: Section 2 describes the data set including catalogue information. The theory behind the  convolutional neural networks and its training is described in section 3. Section 4 discusses the pre-processing and augmentation techniques applied to the original images.  Section 5 focuses on modeling and evaluation of experiments and Section 6 discusses and summarises the overall findings.
\section{Data Sample Selection}
This section describes the data sample selection for this work. The input for our experiments is a dataset of manually annotated images from FIRST catalogue (Faint Images of the Radio Sky at Twenty Centimeters), a survey carried out by Very Large Array \citep{first}. We used different samples for FRI, FRII, Bent-tailed, and compact radio galaxies.

The FRI sample was selected from the FRICAT catalogue  of FRI radio galaxies constructed by \citet{Capetti2017a}. The FRICAT catalogue was compiled by combining observations from NRAO Very Large Array NVSS \citep{Condon1998}, FIRST, and SDSS \citep{York2000} surveys. This catalogue contains sources with redshift $\leq  0.5$ and and edge-darkened radio morphology extending to a radius larger than $30\, kpc$ from the centre of the host galaxy. The catalogue contains 219 FRI radio galaxies.

%We also used more FRI sources from the combined NVSS and FIRST galaxies sample (CoNFIG hereafter) compiled by \citet{Gendre2009} and \citet{Gendre2010}.  The CoNFIG sample contains bright radio sources with $S_{1.4GHz}> 1.3\, Jy$.

For FRII, we have made use of the FRIICAT catalogue built by \citet{Capetti2017b}. The FRIICAT catalogue was also constructed by combining observations from FIRST, NVSS, and SDSS surveys. FRIICAT radio galaxies included in this catalogue have an edge-brightened radio morphologies with redshift $\leq 0.5$. The FRIICAT catalogue contains 122 radio sources.

The FRICAT and FRIICAT catalogues are subset of the sample of 18286 radio sources built by \citet{Best2012}. \citet{Capetti2017a, Capetti2017b} visually inspected the FIRST images and preserved sources with edge-darkened (FRI) or edge-brightened (FRII) morphologies  whose radio emission reaches a distance of at least 30kpc from the centre of the host galaxy. This classification has been performed independently by the three authors, and only included the sources for which FRI or FRII classification is proposed by at least two of the three authors.

Bent radio galaxies have been selected from the catalogue compiled by \citet{Proctor2011}. Sources in this catalogue were selected from FIRST survey (2003 April release, 811,117 sources). \citet{Proctor2011} sorted and classified FIRST sources into singles, doubles, and triples groups. This author identified a total number of 7106 groups with four or more component and classified them as bent types (hereafter BENT) including Wide-Angle Tail (WAT), Narrow-Angle Tail (NAT), ring, double-double, X-shaped, hybrid morphology, giant radio sources, etc. \citet{Proctor2011} generated different tables for different morphologies (for example, sources classified as WATs and NATs are included in Table 1 of \citet{Proctor2011}, W-shaped sources included in Table 2, Ring type sources included in Table 5, etc.). For this work, we used only sources identified as WATs and NATs in Table 1 that was compiled by \citet{Proctor2011}. This table also includes sources with uncertain classification, and these sources are marked by "?" next to the classification in the table. Our BENT-tailed sample contains only the confirmed WATs and NATs radio galaxies which amount to 196 sources. Sources with uncertain identification were excluded.
\begin{figure}
    \centering
    \includegraphics[width=\columnwidth,angle=0]{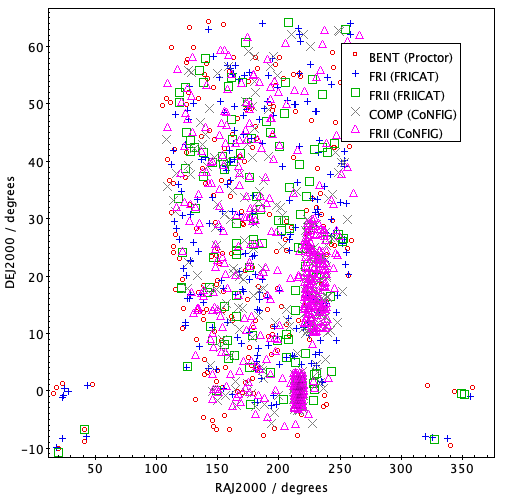}
    \caption{Distribution of the different samples. FRIs from FRICAT (blue plus signs squares), FRII from FRIICAT (green squares), FRIIs from CoNFIG (purple triangle) COMPT sources from CoNFIG catalogue (grey crosses), and Bent-tailed sources from the catalogue compiled by \citet{Proctor2011} (red squares).}
    \label{fig:Map_dat}
\end{figure}

Lastly, the compact radio sources were selected from the combined NVSS-FIRST Galaxy catalogue (CoNFIG) compiled by \citet{Gendre2008} and \citet{Gendre2010}. The CoNFIG catalogue includes new VLA observations, optical identifications and redshifts estimate of compact, FRI, and FRII radio galaxies.This catalogue includes 859 sources over four samples (CoNFIG-1, -2, -3 and -4 with flux density limits of $ S_{1.4 GHz} = 1.3, 0.8, 0.2,$ and $0.05\,$Jy, respectively). The catalogue consists of 71 FRIs radio galaxies, 466 FRIIs radio galaxies and 285 Compact radio galaxies.  We also made use of FRIIs sources in this catalogue and added them to the FRIICAT sources \cite[compiled by ][]{Capetti2017b}. Figure ~\ref{fig:Map_dat} shows the distribution of the different original samples (FRI, FRII, COMPT and BENT).

As reported in \citet{3} and \citet{Ma2019}, we filtered the preliminary samples by excluding sources with uncertain morphologies. By visual inspection, we have excluded sources with strong artifacts. We have also removed Compact and Bent-tailed sources from FRICAT and FRIICAT catalogues, too extended sources, too large sources (to fit in the cutout),  and very small images (insufficient structural information). The resulting sample consists of 125 FRIs, 214 FRIIs, 104 Bent-tailed and 83 compact radio galaxies (see Table ~\ref{tab:sample_selection}). \citet{2} selected their FRIs from FRICAT catalogue, FRIIs from FRIICAT and CoNFIG catalogues, and COMPT sources were selected from CoNFIG. In \citet{3}, FRIs were selected from FRICAT and CoNFIG catalogues, and FRII were selected from CoNFIG catalogue. Both authors selected their Bent-tailed samples from the catalogue compiled by \citet{Proctor2011}. Table ~\ref{tab:sample_selection} summarises the final total number of sources from each class used by \citet {3} and \citet{2} compared to the number of sources used in this work.  The total number of sources used in this work is summarised in Table ~\ref{tab:dataset}.

\begin{table}
\caption{Number of sources and catalogues used in this work compared to previous works. Sources identified as COMPT or Bent-tailed in FRICAT or FRIICAT are not included in this work. Sources marked as unconfirmed or uncertain in CoNFIG catalogue were also removed.}
    \centering
    \begin{tabular}{l|l|c}
    \hline
  Type   & Catalogue  & Source Number  \\
     \hline
     \citet{3} &    & \\
         \hline
    BENT     &   \citet{Proctor2011}   &    254 \\
    COMPT     &   -  &   - \\
    FRIs     &   CoNFIG + FRICAT &   178\\
    FRIIs    &    CoNFIG & 284 \\

      \hline
      \citet{2}   &   &  \\
         \hline
    BENT     &  \citet{Proctor2011}   &   177  \\
    COMPT     &  CoNFIG   &  121 \\
    FRIs     &  FRICAT   &  201 \\
    FRIIs    &   FRIICAT + CoNFIG  & 338 \\
    
      \hline
      This work   &   &  \\
         \hline
    BENT     &  \citet{Proctor2011}    &   104  \\
    COMPT     &  CoNFIG   &  83  \\
    FRIs     &   FRICAT  &  125 \\
    FRIIs    &  FRIICAT + CoNFIG   & 214\\
    
    \hline
    \end{tabular}
    \label{tab:sample_selection}
\end{table}

\section{Convolutional Neural Networks}
Convolutional Neural Network (CNN, \citet{cnn}) is a special form of deep neural networks, which often has an imaging data as an input and can extract necessary features from them for later classification of objects. It is the main difference to the traditional paradigm of machine learning, where features are usually extracted manually. In the case of deep learning models, like convolutional neural networks, the training process will extract and code necessary features automatically without great pre-processing effort. The basic CNN is a combination of different layers in the feed-forward multi-layer architecture. The particular layers extract different levels of patterns. While in the first layers, patterns are more simple shapes like edges, contours, in more deep layers, we can find more complex patterns. The main advantage of CNN architecture, especially for imaging data, is its ability to extract spatial patterns in images\footnote{For more information on deep learning and convolutional neural networks, we recommend \cite{Goodfellow-et-al-2016}.}. 

CNN usually contains several convolutional and pooling layers, which are responsible for the extraction of features and are often followed by fully connected layers for final classification or prediction. \textit{Convolutional layer} is represented by the set of filters (kernels) applied throughout the image or its representation from the previous layer. The basic operation here is convolution, computed by moving filter (window) in the horizontal and vertical directions, followed by the application of activation function. The mathematical equation for convolution is defined as follows:
     \begin{eqnarray}
     S(i,j)=(K*I)(i,j)=\sum_m \sum_n I(m,n)\;K(i-m,j-n),
    \end{eqnarray}
where $m$ is the number of rows and $n$ number of columns for $I$ as the input image (for the first convolutional layer) or its deeper representation from the previous layer, $K$ is a kernel (filter), and $S(i,j)$ represents the result of convolution for $i$-th row and $j$-th column. The convolutional layer is followed by the \textit{Pooling layer} \citep{maxp}, which transforms the local region of image representation (vector or submatrix) into a scalar value that represents the value of pixels from this region. This operation reduces the number of parameters for the next layers. For the pooling operation, there are usually two options – average or max pooling. 

%In our models, we used max pooling, which selects a maximal value from the local region (window) as its representative value.

As an activation function, there are several options that can be used like \textit{ReLU}, \textit{tanh}, \textit{sigmoid} \citep{Goodfellow-et-al-2016}. For example, one can apply nonlinear function \textit{ReLU} \citep{Nair2010} to hidden and output layers, which is often used in CNN architectures. This function is defined on $(0,\infty)$ as $f(x)=\max(0,x)$. 

At the end of the architecture, the output layer provides the final classification or prediction result. The learning process tries to minimize errors of predictions by changing parameters. In the case of multi-class classification, the loss function often used for error minimization in the training step is known as Cross-Entropy (CE), which is defined as follows:

    \begin{eqnarray}
       CE = -\sum_i^M y_i\;\log(\hat{y}_i),
    \end{eqnarray}
    
where $y$ is real class value (like COMPT, BENT, FRI, FRII), $\hat{y}$ is the predicted class value and $M$ is the number of classes (4 in our case). Output \textit{fully connected layer} uses \textit{Softmax} function, which is an extended logistic function. It has the same interval of output values $(0,1)$ but is used for multi-class classification. The number of neurons is identical to the number of classes, and the \textit{Softmax} activation function normalizes values from input to the vector of output probabilities for particular classes (sum is 1). Therefore, \textit{Softmax} is a categorical probability distribution function and can be mathematically defined as follows:
 \begin{eqnarray}
     f(x_k)= \frac{\mathrm{e}^{x_k}}{\sum_{i=1}^K\mathrm\,{e}^{x_i}},
 \end{eqnarray}
where $k$ indexes $K$ input values, $k=1,2,\dots, K$ \citep{Goodfellow-et-al-2016}.

An essential part of the neural network learning process is the optimization algorithm. Often used optimization function is Adam \citep{adam} – algorithm for adaptive moment optimization, which combines Momentum heuristic \citep{momentum} and Root Mean Square Propagation (RMSProp, \cite{hinton2012neural}). RMSProp and Momentum are two different approaches for gradient search, which can be adaptively combined to achieve faster and more robust optimization. Momentum heuristic accelerates search in the direction of minimums, and RMSProp slows search in the direction of oscillations.

Another aspect of the learning process is regularization, which represents different methods that restrain the learning process and help to avoid overfitting. A very efficient regularization function is  known  as \textit{Dropout} \citep{dropout}. The difference between Dropout and other penalization methods (like L1 and L2 norm \citep{ng2004feature}) is that it will not reduce model complexity, but it will effectively reduce the variance of learning. This method is often used for regularization in deep learning, and it is based on the random selection of neurons in every iteration and their removal from current computation, including their connections to the previous and next layer.
\subsection{Metrics for evaluation of classification results}
There are several classification metrics that can be used to evaluate the performance of a classifier. The use of various metrics is important to understand robustness of a classifier and to avoid over-fitting. The most common way of quantifying the classification relies on the concepts of recall and precision \citep{Ivezic2014}. Precision refers to the fraction of true positive returned among all returned positive instances, recall is the fraction rate of true positives that get predicted correctly of all the positives in the datasets. The F1 score is the weighted harmonic mean of precision and recall. The accuracy is the total proportion of correct predictions. However, in order to evaluate the accuracy of the model developed in this work, the  precision, recall, F1 score and accuracy  were calculated as given below :

%\begin{align}
\begin{eqnarray}
\label{eq:x}
\mathrm{Precision} = \frac{\mathrm{TP}}{\mathrm{TP+FP}} \\
\mathrm{Recall} = \frac{\mathrm{TP}}{\mathrm{TP+FN}} \\
\mathrm{F1\ score} = \frac{2 \times \mathrm{Precision}\times \mathrm{Recall}}{\mathrm{Precision+ Recall}} \\ 
\label{eq:x2}
\mathrm{Accuracy} = \frac{\mathrm{TP}+\mathrm{TN}}{\mathrm{TP+FP+TN+FN}} ,
\end{eqnarray}
%\end{align}

 \noindent where TP refers to the true positives, FP refers to the false positives, and FN refers to false negatives.  For example, if the class for evaluation is BENT. Then TP samples are those for which we know that are BENT and also model predicts BENT. TN samples are predicted as not BENT, and they are not BENT. FN  samples are predicted not to be BENT but are BENT. FP  samples are predicted as BENT but are not BENT.
 
Recall and precision are often used to better understand sensitivity and positive predictivity of the model. Most classification systems try to find some compromise between precision and recall, because they are often in antinomic relation – more effort in increase of precision will decrease recall, and vice-versa. F1 score combines both metrics into one scalar value and provides such compromise for evaluation of classifier \citep{han2011data}.

\section{Image pre-processing and data augmentation techniques}

\subsection{Pre-processing of original images and rotation-based augmentation}

\begin{figure}
    \centering
    \includegraphics[width=\columnwidth,angle=0]{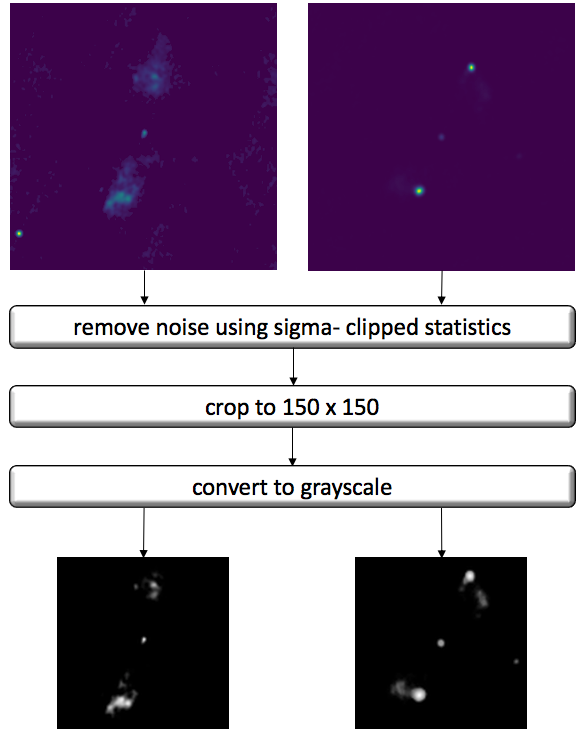}
    \caption{Pre-processing of original images from FIRST catalog using noise reduction based on sigma-clipping, cropping and grayscale transformation.}
    \label{fig:sigma}
\end{figure}

The original dataset had 104 images from BENT class, 83 images from COMPT class, 125 images from  FRI class, and 214 images from FRII class. For the cleaning process we used the same method adopted by \citet{3}, who removed all pixel values lower than a $3\,\sigma$ level of the background and set it to zero. The authors reported that the background noise decreases classifier performance. Similar studies by \citet{2} and \citet{1} followed \citet{3} approach. However, the first pre-processing phase of original images consisted of three steps (see Figure~\ref{fig:sigma}). In the first step we used noise reduction technique based on the filter \texttt{sigma\_clipped\_stats}\footnote{docs.astropy.org/en/stable/api/astropy.stats.sigma\_clipped\_stats.html}, which calculates sigma-clipped statistics on the provided images. The result of this operation is a set of statistics - mean, median, and standard deviation. Then we applied it to images and removed signals from images for values less than a product of $\sigma=3$ and extracted standard deviation. Because galaxy sources are usually in the center of images, we cropped them from the original size $300\times300$ to $150\times150$. Removed pixels that did not bring any significant features to learning. Then the image was transformed into a grayscale format. 

\begin{figure}
    \centering
    \includegraphics[width=\columnwidth,angle=0]{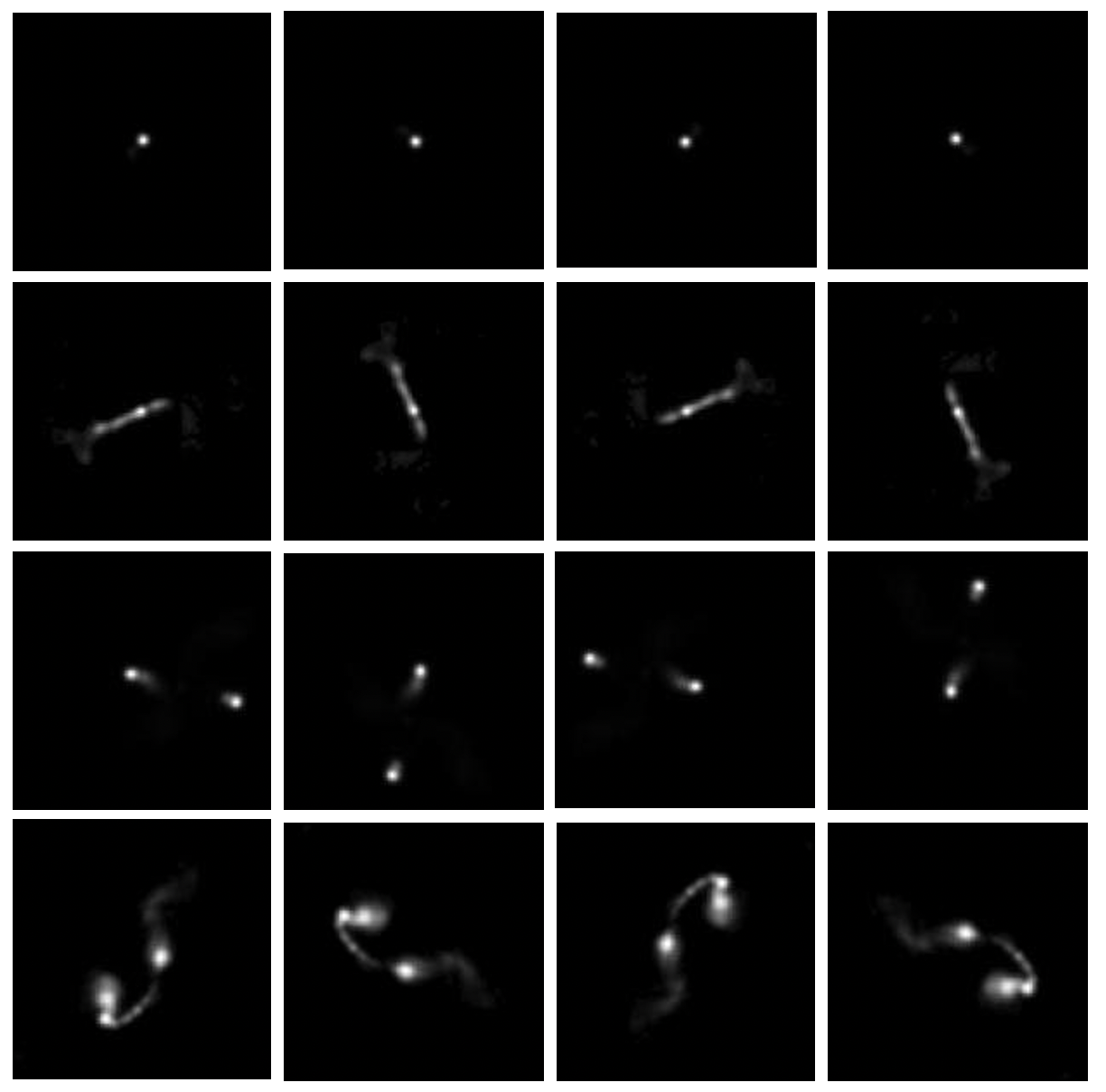}
    \caption{Example of radio galaxies for particular classes after second phase of pre-processing based data augmentation with rotations. First row shows compact radio galaxy, second and third row provide examples of FR type I and FR type II, and the last row is an image of BENT radio galaxy. First image in every row is original image, other images are rotated versions for 90, 180, and 270 degrees.}
    \label{fig:rotate}
\end{figure}

In the second pre-processing phase, we applied rotation to images, for 90, 180, and 270 degrees, which is a simple way of data augmentation often used in the classification of images, if it will not change actual class. Figure~\ref{fig:rotate} shows some rotated images, where every row provides examples for different classes (from top to bottom: COMPT, FRI, FRII, BENT). We also resized images to  $64\times64$ pixels (using re-sampling) in order to reduce the computational needs of the neural network.

We created two testing subsets for the testing phase, following two setup approaches used in previous works by other authors. As was reported by \cite{2}, we rotated all the images and then divided them into a training and testing subset. The split of the extended pre-processed dataset (with rotated images) for training and testing samples was 80:20\%. The number of images in particular classes in original or extended (rotated) versions, as well as sizes of training and testing subsets, are summarised in the first part of Table~\ref{tab:dataset}. We will refer to the testing subset created from the rotated images dataset as the \textit{test set 1}.

In the second part of the experiments, we created a testing subset as described in \cite{3, Lukic_2018, 1}. At first, we divided the original dataset into training and testing subsets in a ratio of 80:20\%. Then we applied the rotations (after the split) only for the images in the training subset. The number of images in particular classes in the original dataset, testing subsets as well as sizes of training subset and their extended (rotated) versions are available in the second part of Table~\ref{tab:dataset}. We will refer to this testing subset as the \textit{test set 2}. 

We also used a subpart of the training set for validation purposes during the training phase. In all cases, validation subsets were 10\% of the final training set. In both setups, training samples also became an input for applying other augmentation techniques, which enhanced the input dataset further.

%Before the next steps we divided extended pre-processed dataset (with rotated images) into training and testing subsets (80:20\%). The number of images in particular classes in original or extended (rotated) version, as well as sizes of training and testing subsets, are summarised in Table~\ref{tab:dataset}.

\begin{table}
	\centering
	\caption{The first part of the table shows the number of original sample images, their number after rotation, and the size of the training and testing subset (referred to as the test set 1) created by the split of already rotated images. The second part of the table shows the number of samples for the second approach, where we split the original dataset into training and testing subsets before applying rotation. In this case, only the training subset then contains rotated images, and we refer to the testing subset as the test set 2.}
	\label{tab:dataset}
	\begin{tabular}{lccccr}
		\hline
		type & original dataset & rotate & test set 1 & train \\
		\hline
		BENT & 104 & 416  & 84 & 332 \\
		COMPT & 83 & 332 & 67 &  265 \\
		FRI & 125 & 500  & 100 &  400\\
		FRII & 214 & 856 & 172 & 684 \\
		\hline
		Total & 526  & 2104 & 423 & 1681\\
		\hline
		\hline
		type & original dataset & test set 2 & train & rotate \\
		\hline
		BENT & 104 & 21 & 83  & 332\\
		COMPT & 83 & 17 &  66  & 264\\
		FRI & 125 & 25 &  100 & 400 \\
		FRII & 214 & 44 & 170 &  680\\
		\hline
		Total & 526  & 107 & 421 & 1676\\
		\hline
	\end{tabular}
\end{table}

\subsection{Other data augmentation techniques}

One of the simplest ways to avoid overfitting is to have enough large and representative dataset. Data augmentation based on the synthetic extension of training set using a modified version of data, helps generalize models and enhance their robustness \citep{han2011data}. It is only necessary to ensure that process will not bring irrelevant data to learning. In this work, it means that any technique will not change the image in the way that change a class or drop the chance to identify a class correctly. In the first pre-processing phase, we already applied rotation-based augmentation to original images, a technique usually used for this purpose. 

\begin{figure}
    \centering
    \includegraphics[width=0.45\textwidth,angle=0]{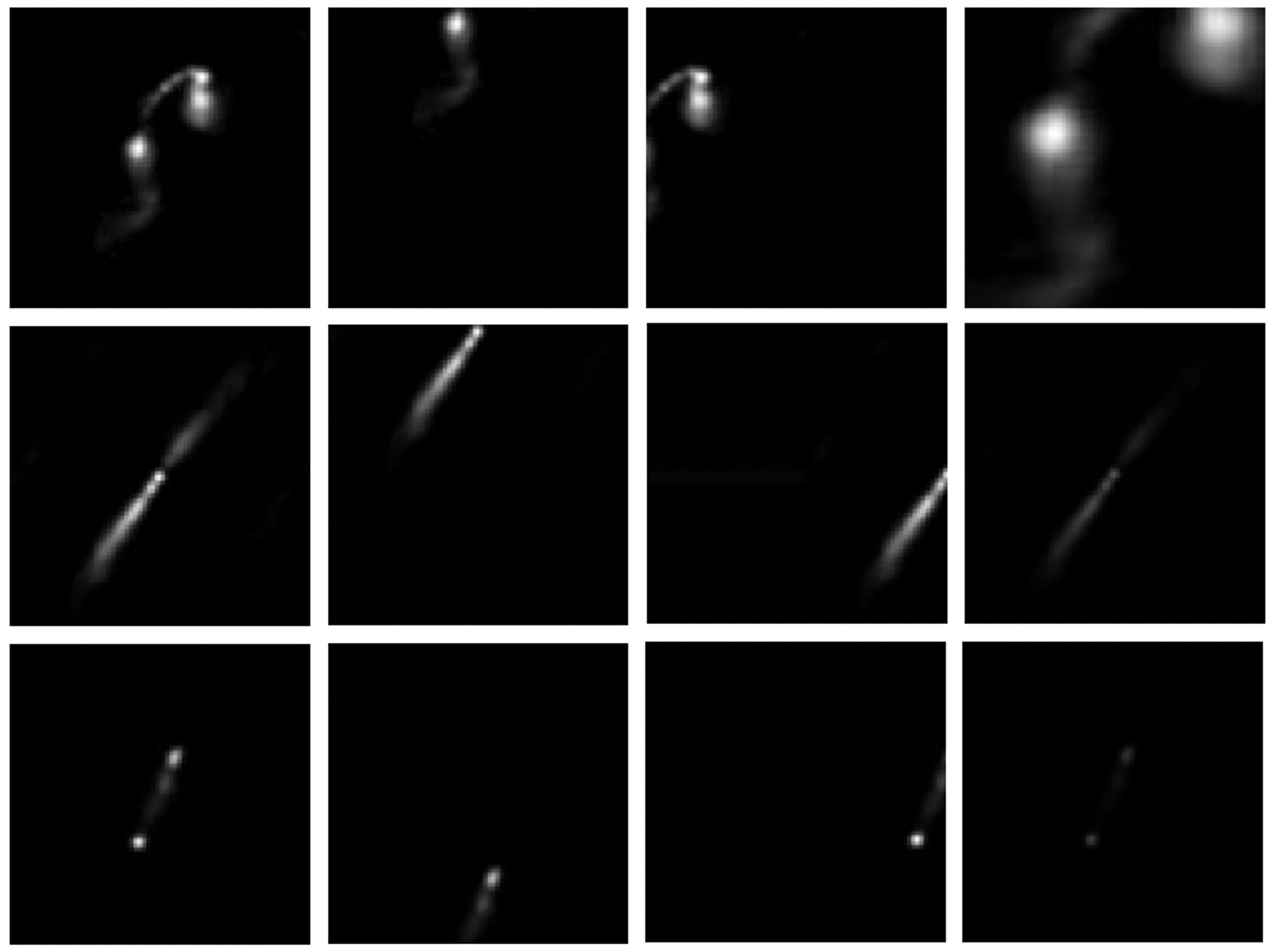}
    \caption{Data augmentations with the negative effect on classifier accuracy -- first column show original images, second height shift, and third width shift. The first image in the last column shows a zoomed image, and the other two images in this column show a decrease of brightness.}
    \label{fig:augmentation}
\end{figure}

Since we started with a small dataset and also that other works \citep{1, 2, 3, Lukic_2018} did not test many augmentations, we also decided to test some other techniques to enlarge our training subset. In this case we used \texttt{ImageDataGenerator}\footnote{https://keras.io/api/preprocessing/image/}, which generates batches of tensor image data with real-time data augmentation. Here it is possible to enhance datasets with additional data augmentations like a vertical or horizontal flip of an image, change of brightness range, zoom, width or height shift. However, we had to be careful here. One obvious example is shifting augmentation. In this case with relatively small images, we could easily bring some problematic samples into the dataset, which would decrease the accuracy of models. Even a small shift would move subpart of extended radio galaxy source out of the boundaries of the image. Similarly, zoom augmentation would displace some parts of the lobes and move them out of the image. Another augmentation with negative effects was decreasing the brightness of the image, which also reduces potential for successful differentiation of images. A visual example of the effects of such improper augmentations is shown in Figure \ref{fig:augmentation}.    

In the first experiment, we tested different combinations of augmentation techniques,which gave us a good overview for the final selection of augmentations. Besides the expected use of rotations with approximately the 6-8\% improvement in accuracy for the original dataset, we were able to quantify the effects of additional augmentations (at least in a relative manner), which then helped us to select the final setup of augmentation techniques. From a quantitative evaluation of these experiments, we knew that the use of a large range of brightness (especially if there is also a reduction in brightness) with settings range [0.35-1.5] worsened the average accuracy of the model by 3\%. Random shifts (vertical or horizontal) with settings range from 0\% to a maximum of 25\% have resulted in an average reduction of the accuracy of 6\%. The application of the zoom function worsened the model by up to 12\%, mainly because the input images were centred and cropped. In general, any combination of techniques -- brightness reduction, zoom, shift with other techniques such as flip, harms the accuracy of the model. The addition of images with increased brightness compared to the base model resulted in an improvement of approximately 3\%.  The combination of vertical and horizontal flip can add 9-11\% to accuracy. The mix of increased brightness levels and flip techniques showed the best results according to the accuracy of the model. Therefore, we used them also for the pre-processing of the training set for the learning of the final version of CNN architecture. The setup of ImageDataGenerator was defined as follows:

\begin{verbatim}
ImageDataGenerator(brightness_range=[1,1.5])
ImageDataGenerator(horizontal_flip=True,
                   vertical_flip=True)
\end{verbatim}

The brightness range from 1 to 1.5 represents brightness from the level of the original image (for 1) to a 50\% increase over the original image (for 1.5).

\begin{figure}
    \centering
    \includegraphics[width=0.45\textwidth,angle=0]{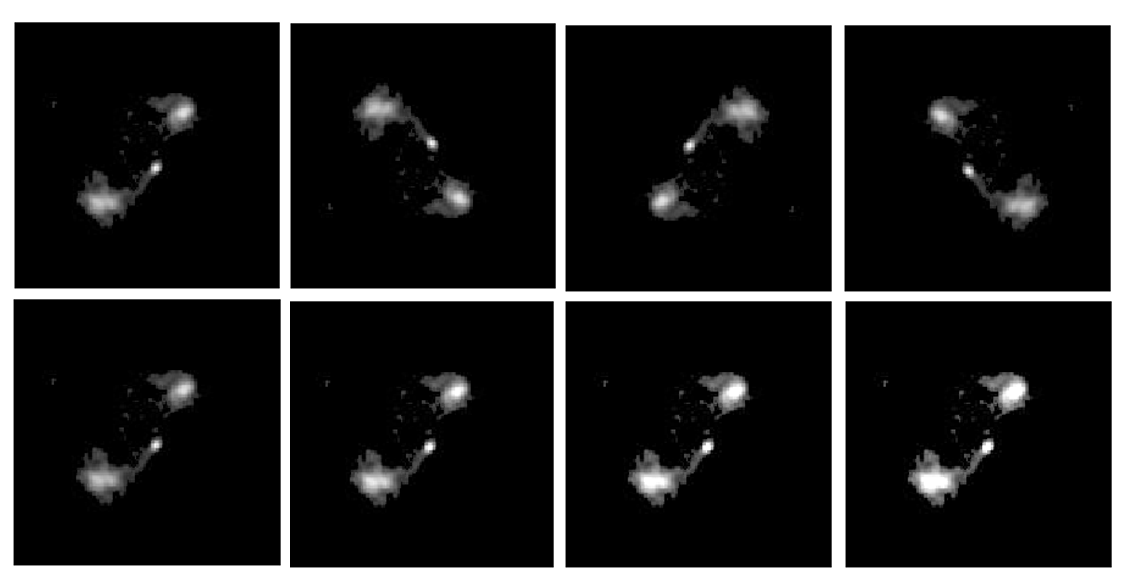}
    \caption{Example of augmentation techniques applied for training of the final classifier. The first row shows the use of \texttt{ImageDataGenerator} with a vertical and horizontal flip. The second row of images shows an incremental increase of brightness, from the original image (brightness level 1) on the left to 50\% more brightness on the right (brightness level 1.5).}
    \label{fig:augmentation2}
\end{figure}

Figure~\ref{fig:augmentation2} shows examples of FRII radio galaxy sources, where the first row shows horizontal and vertical flips, and in the second row are different levels of increased brightness. We applied a technique during the training, which we called a double train classifier. First version of classification model was based on a dataset extended by brightness enhancement augmentation, followed by the model's additional training on the dataset (with already rotated images) with horizontal and vertical flips.  

\section{Modeling and evaluation}\label{eval}

\begin{figure*}
    \centering
     \resizebox{14.7cm}{!}{\includegraphics[angle=90]{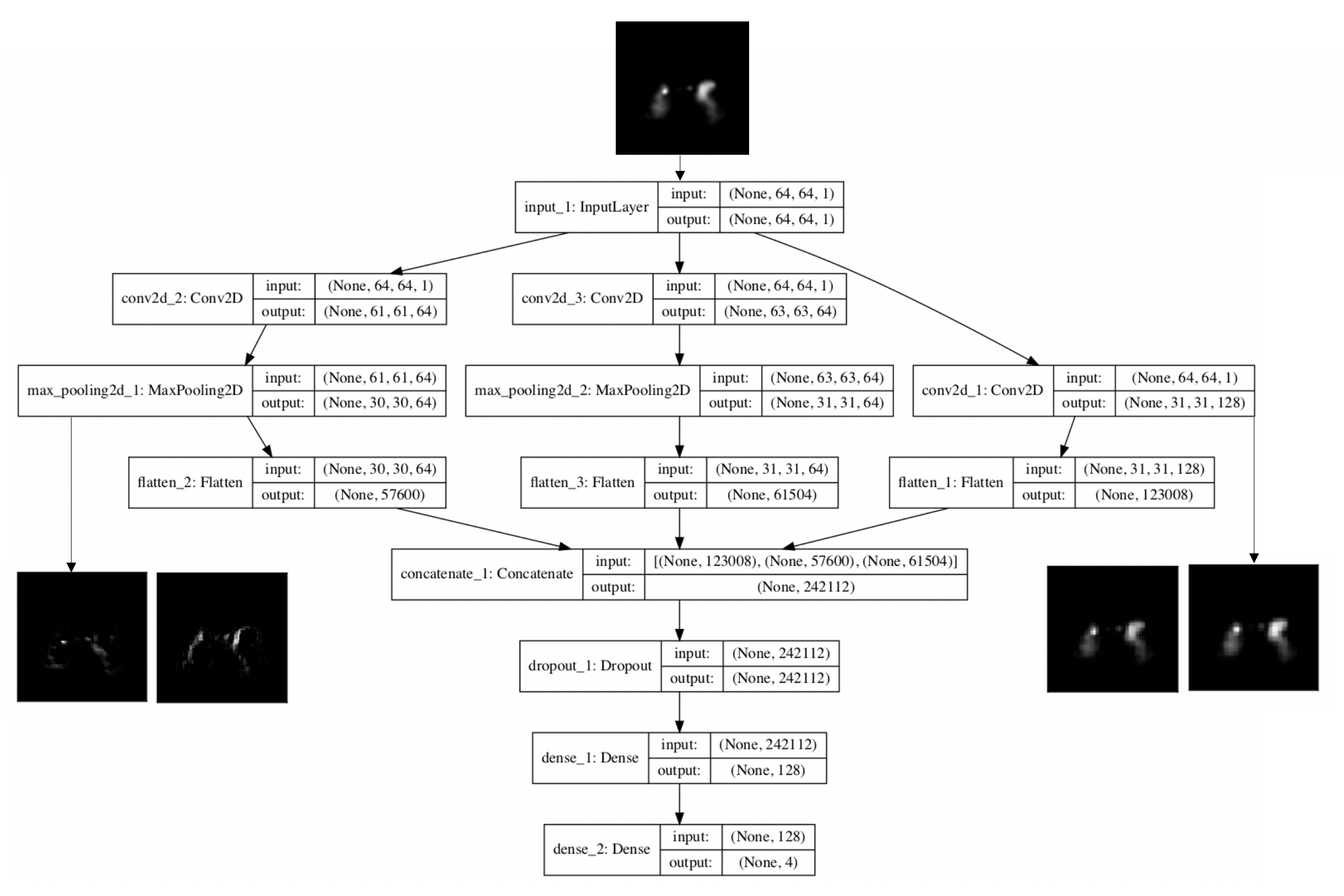}}
    \caption{The architecture of the convolutional neural network with visualization of feature maps. This graph shows particular layers with input and output sizes. An Input image is processed using three convolutional blocks with different filters. The output of convolutions is then combined into one vector in the concatenate layer. After regularization based on dropout, we apply two fully connected (dense) layers for final classification. For convolutional blocks and first dense layer activation function is ReLU. The last layer is the output layer of the architecture with four neurons and Softmax activation function, which represents the classification of input images into four classes.}
    \label{fig:model}
\end{figure*}

In general, for our CNN architecture, we decided to apply the approach of combining three different parallel convolutional blocks (with different basic setups) into one feature vector, which then moves to a fully connected part for final classification. This combined approach helps the classifier to enhance its granularity and robustness. It is based on the idea of inception models from Google for more complex networks, introduced in \cite{szegedy2015going}, but we adapted it for our smaller network. As the authors show in their paper, such an approach based on inception modules allows the full model to process visual information at various scales and then aggregate it for the next layer. Therefore, the abstract features from different scales are available simultaneously and provide better granularity and robustness for the model. Figure~\ref{fig:model} shows the structure of the whole architecture, which we now describe in more detail.    

The first layer of the architecture is the perceptive input layer, where we provide an input image with a size $64\times64$ pixels. This input image is then used as the input of three convolutional neural network blocks. First block (first from right in Figure~\ref{fig:model}) consists of convolutional layer with 128 convolutional filters, window size $3\times3$ and step 2. The second block contains a convolutional layer with 64 filters, $4\times4$ kernel size, and is followed by the max-pooling layer with the $2\times2$ window size. The third convolutional block contains a convolutional layer with 64 filters and kernel size $2\times2$, and max-pooling layer with the $2\times2$ window size. The step size in the second and third convolution layers is one in the horizontal direction and one in the vertical direction. The output of all blocks is transformed into a vector using flatten layers and combined into one vector in the concatenate layer. Then we applied the dropout regularization technique to avoid overfitting. The use of dropout reduces learning variance by a random selection of neurons (including its connections) during every iteration and removes them from the computation. Therefore, every iteration has a slightly different structure of connections. The synergetic effect of this approach  through the learning process can lead to lower variance of the final classifier. Our model is relatively simple, and we expected that droupout rate should be relatively low. We also tested the influence of different dropout rates in the hyperparameters tuning phase, but the expected best setup for dropout was around 20\%. The next part of the architecture consists of two layers with the full connection. First of these layers is  fully connected (dense) layer that reduces feature vector to 128 units. For every layer, until now, we used ReLU as an activation function. The last fully connected layer is the output layer, which uses the Softmax activation function (with 4 units) for final classification into four classes – BENT, COMPT, FRI a FRII. A table-based summary of the architecture with the shape of all layers, the number of trainable parameters, and connection to previous layers are also available in Table~\ref{tab:sum}.

Even if we introduced block-based architecture, it is a relatively small and effective network, also thanks to cropping and resize of input images. Therefore, training and application of the model are fast. The time required to train one epoch is approximately 30 seconds. For the final prediction, we need only one second for the whole test set. For clarification, within this paper, we refer to an epoch as a term from artificial neural networks – it is one training cycle through the full training dataset. For the training of the final classifier, we used hyper-parameters depicted in Table~\ref{tab:hyperparam}.

\begin{table}
\centering
\caption{Summary of the architecture of our final CNN model in table form. We can see all the layers with their shape, trainable parameters, and previous layer(s) connections. At the end of the table is a summary of trainable parameters for the whole network.}
\begin{tabular}{llll}
\hline                                    
\textbf{Layer (type)  }                                & \textbf{Output Shape }  & \textbf{\textbf{Params.}}  & \textbf{Previous layer}                                              \\ \hline
InputLayer & (None, 64, 64, 1) & 0  &         \\
Conv2D\_1 & (None, 31, 31, 128) & 1 280 & InputLayer   \\
Conv2D\_2 & (None, 61, 61, 64) & 1 088  & InputLayer       \\
Conv2D\_3  & (None, 63, 63, 64) & 320 & InputLayer   \\
MaxPool2D\_1 & (None, 30, 30, 64)   & 0 & Conv2D\_2          \\
MaxPool2D\_2  & (None, 31, 31, 64)     & 0 & Conv2D\_3   \\ 
Flatten\_1 & (None, 123 008)     & 0 & Conv2D\_1   \\
Flatten\_2 & (None, 57 600)     & 0  & MaxPool2D\_1 \\
Flatten\_3 & (None, 61 504)     & 0  & MaxPool2D\_2 \\
Concatenate & (None, 242 112)     & 0 &  Flatten\_1 \\
 &     &  &  Flatten\_2 \\
 &     &  &  Flatten\_3 \\
Dropout  & (None, 242 112)     & 0 & Concatenate   \\
Dense\_1 & (None, 128)     & 30 990 464 & Dropout    \\
Dense\_2  & (None, 4)     & 516  & Dense\_1   \\
\hline
\textbf{Total params:}                   &   30 993 668              &          \\
\textbf{Trainable:}                    &    30 993 668           &          \\
\textbf{Non-trainable:}                     &     0          &         \\ \hline
\end{tabular}
\label{tab:sum}
\end{table}

\begin{table}
	\centering
	\caption{Hyper-parameters used to train our final classifier.}
	\label{tab:hyperparam}
	\begin{tabular}{cc} % four columns, alignment for each
		\hline
		 \textbf{Hyper-parameters} & \textbf{Values}\\
		\hline
	     Epochs & 50\\
         Batch Size & 32\\
		 Learning rate & $0.001$\\
		 Dropout rate &  20\% \\
		\hline
	\end{tabular}
\end{table}

We trained our neural network architecture first with the extended dataset (containing rotated images) using brightness range augmentation for the range $(1,1.5)$. Then we additionally trained the same models on dataset augmented by the horizontal and vertical flip. This technique we  called  a double train classifier and is inspired by the transfer learning approach. It means that the first training provides pre-trained weights of a network similar to a neural network used in transfer learning for additional adaptation to a new domain. A similar approach inspired by transfer learning was successfully used in \cite{1}, where the authors used weights of pre-trained models from FIRST catalog for classification of images in NVSS catalog and vice-versa.

Neural networks are often described as black-box models. Convolutional neural networks provide some interpretability aspects of their architecture. One of them is the visualization of feature maps. The feature map is the output of one filter applied to the previous layer and it shows how this filter extract and generalize features from training samples. It means that we can show image-like visualizations generated by applying filters or feature detectors to the output of a specific layer. Such visualization helped us to understand better what the network learned qualitatively and identify early some problems which lead to the wrong prediction. It can be helpful during the network-tuning phase. Therefore, during the modeling phase, we also watched and visually evaluated feature maps extracted from different layers within the architecture. In Figure~\ref{fig:model}, we can see not only the architecture of our model but also some examples of feature maps. Using these feature maps, we can also see how the block-based architecture extracts details on different scales or granularity.

\begin{figure}
    \centering
    \includegraphics[width=0.495\columnwidth]{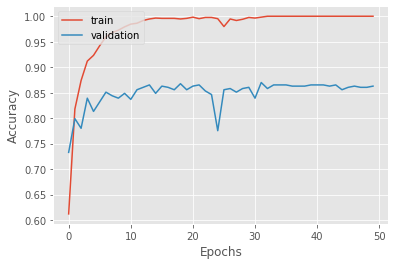}
    \includegraphics[width=0.495\columnwidth]{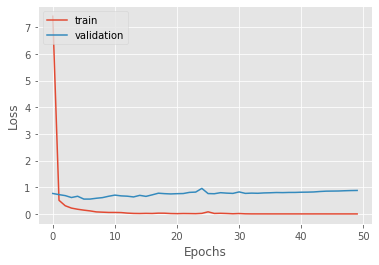}
     \includegraphics[width=0.495\columnwidth]{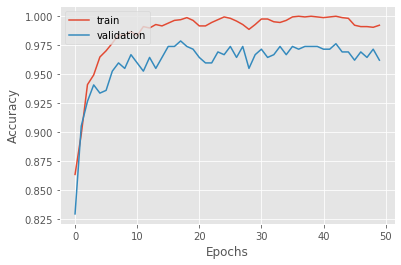}
    \includegraphics[width=0.495\columnwidth]{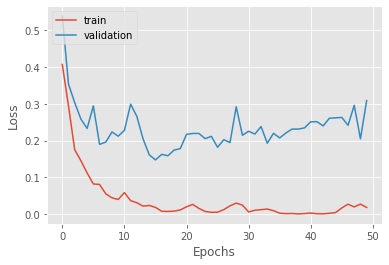}
    \caption{Effect of double train classifier. Learning curves showing the accuracy and loss of our CNN model in individual epochs for the sequence of training phases for different augmentations. The first line shows the learning process graphs after the first training and the second line after the second training. We can see the higher accuracy and lower loss at the start of training in the second row.}
    \label{fig:Acc}
\end{figure}

After fixing the final version of hyper-parameters, we started the training and evaluation process. An example of the progress of the training is shown in Figure \ref{fig:Acc}. The first row shows our CNN model's learning curves in individual epochs on the training set and their respective accuracy and loss after the first training. The second row provides learning curves after the second training with additional augmentations. We can see the effect of training in sequence on the curve in the second row, where the pre-trained model from the first training phase helped to start at much higher values of accuracy and much lower values of loss. To ensure the robustness of models, we trained and tested our architecture for 5 different splits of training and testing samples and also for different augmentations generated using ImageDataGenerator. We also tested particular splits of the dataset in more runs (5-10), and the variance of learning showed that uncertainty of the results was under 1\%. We evaluated accuracy, precision, recall, and F1 score for every predicted class. Also, we calculated averaged values of evaluation metrics for the whole dataset using a micro averaging approach – it is a method averaging metrics based on the number of all samples evaluated as true positives, false negatives, and false positives. For the evaluation of the model we used classification report from \texttt{sklearn.metrics} package.

 The evaluation results  on test set 1 (with rotation) are shown in Table~\ref{tab:result}. The first part contains the results of classification without additional real-time augmentation techniques (ImageDataGenetator). The last part contains average values of five previous experiments with applied data augmentations. For compact radio galaxy sources (COMPT class), our model achieves 100\% precision, recall, and F1 score. For extended radio galaxy sources (classes FRI, FRII, BENT), our classifier achieves an average 99\% in all evaluated metrics. Generally, our classifier performs with averaged accuracy of 99\% for all classes. 

\begin{table}
	\centering
	\caption{Results: test set 1. First sub-table shows the evaluation results of classifier without additional augmentation techniques. The next five sub-tables contain the model's evaluation results with data augmentations for different random divisions of the dataset to training and testing parts. The last table shows the average values of these experiments trained on the extended dataset. Support represents the number of test samples for each class, with sum of them as total.}
	\label{tab:result}
	\begin{tabular}{lcccr} 
		\hline
	    \textbf{Type} & \textbf{Precision} & \textbf{Recall} & \textbf{F1 score} &  \textbf{Support} \\ 
	    \hline
		\multicolumn{5}{l}{\textbf{Without data augmentation} -- accuracy: 0.89}   \\ \hline
		BENT & 0.82 & 0.79 &  0.80 & 84\\
		COMPT & 0.96 & 1.00 &  0.98  &67 \\
		FRI & 0.85 &  0.99&  0.90 & 100\\
		FRII & 0.89 &0.83 & 0.86 & 172\\
		\hline
		\textbf{avg} & \textbf{0.88}  &\textbf{ 0.88} & \textbf{0.88} & \textbf{total: 423}\\
		\hline \hline
		\multicolumn{5}{l}{\textbf{Data augmentation -- 1. split} -- accuracy: 0.99}   \\ \hline
		BENT & 1.00 & 0.99 &  0.99 & 84\\
		COMPT & 0.99 & 1.00 &  0.99  &67 \\
		FRI & 0.99 &  0.99&  0.99 & 100\\
		FRII & 0.99 &0.99  & 0.99 & 172\\
		\hline
		\textbf{avg} & \textbf{0.99}  &\textbf{ 0.99} & \textbf{0.99} & \textbf{total: 423}\\
		\hline
		\multicolumn{5}{l}{\textbf{Data augmentation -- 2. split} -- accuracy: 0.99}   \\ \hline
	    BENT & 1.00 & 0.98& 0.99  & 84\\
		COMPT & 0.99 & 1.00 &  0.99  &67 \\
		FRI & 0.99 & 0.99 &  0.99 & 100\\
		FRII & 0.99 &1.00  & 1.00 & 172\\
		\hline
		\textbf{avg} &  \textbf{0.99} & \textbf{0.99} & \textbf{0.99} & \textbf{total: 423}\\
		\hline
		\multicolumn{5}{l}{\textbf{Data augmentation -- 3. split} -- accuracy: 1.00}   \\ \hline
		BENT &1.00  &1.00 & 1.00  & 84\\
		COMPT & 1.00 & 1.00 &   1.00 & 67\\
		FRI & 1.00 & 1.00 &  1.00 & 100\\
		FRII &1.00  &  1.00& 1.00 & 172\\
		\hline
		\textbf{avg} & \textbf{1.00}  &  \textbf{1.00}& \textbf{1.00} & \textbf{total: 423}\\
		\hline
		\multicolumn{5}{l}{\textbf{Data augmentation -- 4. split} -- accuracy: 0.99}   \\ \hline
		BENT & 0.99 & 0.98  & 0.99  &84 \\
		COMPT & 1.00 & 1.00 &   1.00    &67 \\
		FRI & 0.97 & 0.99 &  0.98 & 100\\
		FRII & 0.99 & 0.98 & 0.99 & 172\\
		\hline
		\textbf{avg} & \textbf{0.99}   & \textbf{0.99}  &\textbf{0.99}   & \textbf{total: 423}\\
		\hline
		\multicolumn{5}{l}{\textbf{Data augmentation -- 5. split} -- accuracy: 0.98}   \\ \hline
		BENT & 0.97 & 0.99 & 0.98  &84 \\
		COMPT & 0.99 & 1.00 &   0.99   &67 \\
		FRI &1.00 & 0.97 &   0.98  & 100\\
		FRII & 0.98 & 0.98 &   0.98  &172 \\
		\hline
		\textbf{avg} &  \textbf{0.98} & \textbf{0.98} &   \textbf{0.98}  & \textbf{total: 423}\\
		\hline \hline
	    \multicolumn{5}{l}{\textbf{Data augmentation -- Average} -- accuracy: 0.99 }  \\ \hline
		BENT & 0.99 & 0.99 & 0.99  &84 \\
		COMPT & 1.00 & 1.00 &   1.00   &67 \\
		FRI &0.99 & 0.99 &   0.99  & 100\\
		FRII & 0.99 & 0.99 &   0.99  &172 \\
		\hline
		\textbf{avg} &  \textbf{0.99} & \textbf{0.99} &   \textbf{0.99}  & \textbf{total: 423}\\
		\hline 
	\end{tabular}
\end{table}

 The view on the performance of the model built on an independent testing subset (test set 2) is available in Table \ref{tab:result_test2}. The first sub-table contains the results without using rotation and ImageDataGenerator. The training set, in this case, contained only 421 images. To ensure the robustness of models, we trained and tested our architecture for five different training and testing splits. Augmentation techniques (rotation and ImageDataGenerator) were used only for the training subset. The average results of these experiments are in the last sub-table of Table~\ref{tab:result_test2}. Even in this case, the best model is able to successfully classify both compact and extended radio galaxy sources. The average value of the classification on the test set 2 is 96\% for the precision, recall, and F1 score. Generally, our classifier performs with an average accuracy of 95,4\% for all classes. Table \ref{tab:cm} shows the confusion matrix of the best performance model. We can see that in this case, the model incorrectly classified only three radio galaxy images. 
 
\begin{table}
    \caption{Confusion matrix of the best model performance for test set 2.}
    \label{tab:cm}
    \centering
    \begin{tabular}{@{}llllll@{}}
        \hline
        \multicolumn{1}{l}{}                                  & \multicolumn{5}{c}{\textbf{Actual}} \\ \hline
        \multicolumn{1}{c|}{\multirow{5}{*}{\textbf{Predicted}}} &              & \textbf{BENT}     & \textbf{COMPT} & \textbf{FRI } & \textbf{FRII} \\
        
        \multicolumn{1}{c|}{}   & \textbf{BENT}  & 21       &    0  &  0  & 0 \\
        \multicolumn{1}{c|}{}   & \textbf{COMPT}  &  0     &   17 & 0  & 0\\
        \multicolumn{1}{c|}{}   &  \textbf{FRI} &  0  &0    &  24 &1\\ 
        \multicolumn{1}{c|}{}   &  \textbf{FRII} &  0 &  0  & 2 & 42  \\\hline
    \end{tabular}
\end{table}

\begin{table}
	\centering
	\caption{Results: test set 2. First sub-table shows the evaluation results of the classifier without additional augmentation techniques. The next five sub-tables contain classifier evaluation results using data augmentations with different random divisions of the dataset to training and testing parts. The last table shows the average values of these experiments trained on the extended dataset. Support represents the number of test samples for each class, with sum of them as total.}
	\label{tab:result_test2}
	\begin{tabular}{lcccr} 
		\hline
	    \textbf{Type} & \textbf{Precision} & \textbf{Recall} & \textbf{F1 score} &  \textbf{Support} \\ 
	    \hline
		\multicolumn{5}{l}{\textbf{Without data augmentation} -- accuracy: 0.81}   \\ \hline
		BENT & 0.76 & 0.76 & 0.76 & 21\\
		COMPT & 0.89 & 1.00 &  0.94  & 17\\
		FRI & 0.79 & 0.76 & 0.78  & 25\\
		FRII & 0.81 & 0.80 & 0.80 & 44\\
		\hline
		\textbf{avg} & \textbf{0.82}  &\textbf{ 0.83} & \textbf{0.82} & \textbf{total: 107}\\
		\hline \hline
		\multicolumn{5}{l}{\textbf{Data augmentation -- 1. split} -- accuracy: 0.94}   \\ \hline
		BENT & 0.95 & 1.00 &  0.98 & 21\\
		COMPT & 1.00 & 1.00 &  1.00  &17 \\
		FRI & 0.88 &  0.88 &  0.88  & 25\\
		FRII & 0.95 & 0.93  & 0.94 & 44\\
		\hline
		\textbf{avg} & \textbf{0.95}  &\textbf{ 0.95} & \textbf{0.95} & \textbf{total: 107}\\
		\hline
		\multicolumn{5}{l}{\textbf{Data augmentation -- 2. split} -- accuracy: 0.97}   \\ \hline
		BENT & 1.00 & 1.00 &  1.00 & 21\\
		COMPT & 1.00 & 1.00 &  1.00  &17 \\
		FRI & 0.92 &  0.96 &  0.94 & 25\\
		FRII & 0.98 & 0.95  & 0.97 & 44\\
		\hline
		\textbf{avg} &  \textbf{0.97} & \textbf{0.98} & \textbf{0.98} & \textbf{total: 107}\\
		\hline
		\multicolumn{5}{l}{\textbf{Data augmentation -- 3. split} -- accuracy: 0.97}   \\ \hline
		BENT & 0.95 & 1.00 &  0.98 & 21\\
		COMPT & 0.94 & 1.00 &  0.97  &17 \\
		FRI & 1.00 &  0.92 &  0.96 & 25\\
		FRII & 0.98 &0.98  & 0.98 & 44\\
		\hline
		\textbf{avg} & \textbf{0.97}  &  \textbf{0.97}& \textbf{0.97} & \textbf{total: 107}\\
		\hline
		\multicolumn{5}{l}{\textbf{Data augmentation -- 4. split} -- accuracy: 0.95}   \\ \hline
		BENT & 1.00 & 0.95 &  0.98 & 21\\
		COMPT & 1.00 & 1.00 &  1.00  &17 \\
		FRI & 0.92 &  0.92 &  0.92 & 25\\
		FRII & 0.93 & 0.95  & 0.94 & 44\\
		\hline
		\textbf{avg} & \textbf{0.96}   & \textbf{0.96}  &\textbf{0.96}   & \textbf{total: 107}\\
		\hline
		\multicolumn{5}{l}{\textbf{Data augmentation -- 5. split} -- accuracy: 0.94}   \\ \hline
		BENT & 0.90 & 0.90 &  0.90 & 21\\
		COMPT & 0.97 & 1.00 &  0.99  &17 \\
		FRI & 0.93 &  1.00 &  0.96 & 25\\
		FRII & 0.96 & 0.91  & 0.93 & 44\\
		\hline
		\textbf{avg} &  \textbf{0.94} & \textbf{0.94} &   \textbf{0.94}  & \textbf{total; 107}\\
		\hline \hline
	    \multicolumn{5}{l}{\textbf{Data augmentation -- Average} -- accuracy: 0.95 }  \\ \hline
		BENT & 0.96 & 0.97 & 0.97  & 21\\
		COMPT & 0.98 & 1.00 &  0.99 &17 \\
		FRI & 0.93 & 0.94 &  0.93 & 25\\
		FRII & 0.96 & 0.94  & 0.95 & 44\\
		\hline
		\textbf{avg} &  \textbf{0.96} & \textbf{0.96} &   \textbf{0.96}  & \textbf{total: 107}\\
		\hline 
	\end{tabular}
\end{table}

% Now we would like to discuss the relevancy of the results.
As already mentioned, we tested our classifier using two experimental setups based on the related work. The setup inspired by the work of  \cite{2} increased the dataset by rotation and then divided it into a testing and training subset. In the second case, inspired by the other related work papers, i.e., \cite{3, Lukic_2018, 1}, we separated the test set initially before any augmentation. According to our information on data samples selection, we can see that while original samples of radio galaxy sources are not the same, the difference in their sizes is generally small and are selected similarly from the same catalogs. Also according to fact that we followed both experimental setups, for which we achieved better results, architecture with parallel convolutional blocks combined with additional augmentations showed its ability to improve classification results, even for a slightly smaller number of input images in our case. In an experimental setup with rotation before train-test split results show an improvement of 2\% on average. Consistently with this fact, for the experimental setup with a test set 2 (without rotated version before train-test split), we also achieved on average 2\% better evaluation metrics (in comparison to \cite{Lukic_2018}). In this case, the data selection process is even closer to our (according to number of data sample images) and the independent selection of the test set proves the generalization ability of our model. Therefore, it is evident  that architecture of this type and the effect of selected augmentations can generally improve the classification of radio galaxy sources.

The current work will be extended to include other types of radio galaxy morphologies such as X-shaped and ring-like radio galaxies. We will also work on enhancing the current developed model and its implementation to retrain it for large number of sources that will be collected from the upcoming radio surveys such as SKA \citep{dewdney2009square}, MeerKAT \citep{Jonas2016}, MeerKLASS \citep{Santos2017}, and EMU \citep{Hopkins2015} etc.

All the models were implemented in Python, with Tensorflow \citep{abadi2016tensorflow} and Keras \citep{keras} for neural networks, and Jupyter notebooks are publicly available online at \url{https://github.com/VieraMaslej/RadioGalaxy}. The experiments were conducted on a PC equipped with an Intel Core i7 processor clocked at 2,9 GHz, 16 GB RAM, and GPU Radeon Pro 560 4GB.

\section{Conclusions}
In this paper, we have introduced a classifier based on convolutional neural networks that can automatically classify radio galaxies into four morphological classes - compact radio galaxy sources, and three sub-types of extended radio galaxy sources (Fanaroff-Riley Type I and Type II, Bent-tailed).  Our classifier, based on the combination of three convolutional neural network blocks followed by the two fully connected layers for final classification, showed accuracy, recall, and F1 score results comparable to the state-of-the-art classifiers published in previous works on the morphological classification of radio galaxies. One of the essential aspects was also the augmentation of input data images. We trained the model using images augmented by rotation and added brightness, thanks to which the model was able to better learn morphological properties, especially in extended radio galaxies. The learned weights were reflected in the second training of the model, in which we extended the training set of data with their vertical, horizontal, or both vertically and horizontally flipped copies. We tested two experimental setups based on related work, split of original images (a total of 526 images) before augmentations and selection of testing subset from original images with already applied rotations. Both setups have  shown promising results and achieved better performance to approaches provided in related work, with comparable setups. In the case of a test set 2, we achieved an average of 96\% for precision, recall, and F1-score.  Due to resizing and relatively small input images, the architecture is also computationally effective and can be easily applied for automatic classification tasks.

\section*{Acknowledgements}
This work was supported by Slovak APVV research grant under the contract No. APVV-16-0213 and Slovak VEGA research grant No. 1/0685/21. We thank the anonymous referee for the comments and suggestions that have improved the manuscript considerably.

\section*{Data Availability}
The data underlying this article will be shared on reasonable request to the corresponding author.


\bibliographystyle{mnras}
\bibliography{MNRAS_RG_Final.bib} % if your bibtex file is called example.bib

%\begin{thebibliography}{99}
%\end{thebibliography}

\bsp	% typesetting comment
% Don't change these lines
\label{lastpage}

\end{document}